\documentclass[11pt]{article}

\usepackage{arxiv}

\usepackage[utf8]{inputenc} 
\usepackage[T1]{fontenc}    
\usepackage{hyperref}       
\usepackage{url}            
\usepackage{booktabs}       
\usepackage{amsfonts}       
\usepackage{nicefrac}       
\usepackage{microtype}      
\usepackage{lipsum}
\pdfoutput=1
\usepackage{graphicx}
\usepackage{color}
\usepackage[per-mode=symbol, separate-uncertainty=true]{siunitx}
\usepackage{csquotes}
\usepackage{textcomp}
\usepackage{tikz}
\usetikzlibrary{decorations.pathreplacing}
\usepackage{verbatim}
\usepackage{booktabs}
\usepackage{caption}
\usepackage{amsmath}
\usepackage{colortbl}
\usepackage{subcaption}
\newboolean{color}        
\setboolean{color}{true}  

\ifthenelse{\boolean{color}}
{\definecolor{gold}{HTML}{FFD700}
	\definecolor{rad_yellow}{HTML}{FFFF00}
	\definecolor{radio_green}{HTML}{89AB02}
	\definecolor{copper}{HTML}{B87333}
	\definecolor{crimson}{HTML}{DC143C}
	\definecolor{uni_blue}{HTML}{07529A}
	\definecolor{uni_grau}{HTML}{909085}}
{\definecolor{gold}{HTML}{BBBBBB}	
	\definecolor{rad_yellow}{HTML}{CCCCCC}
	\definecolor{radio_green}{HTML}{89AB02}
	\definecolor{copper}{HTML}{555555}
	\definecolor{crimson}{HTML}{DC143C}
	\definecolor{uni_blue}{HTML}{07529A}
	\definecolor{uni_grau}{HTML}{909085}}

\DeclareSIUnit\hole{hole}
\DeclareSIUnit\e{e}
\DeclareSIUnit\electrons{electrons}
\DeclareSIUnit\ppm{ppm}
\DeclareSIUnit\ppmv{ppmV}

\definecolor{gold}{HTML}{FFD700}
\definecolor{radio_green}{HTML}{89AB02}
\definecolor{copper}{HTML}{B87333}
\definecolor{crimson}{HTML}{DC143C}
\definecolor{uni_blue}{HTML}{07529A}
\definecolor{uni_grau}{HTML}{909085}

\title{Measurements of the Charging-Up Effect in Gas Electron Multipliers}

\newcommand{\rom}[1]{\uppercase\expandafter{\romannumeral #1\relax}}
\newcommand{\trademark}{\raisebox{1mm}{\small{\textregistered}}}

\author{
	Philip Hauer\thanks{hauer@hiskp.uni-bonn.de},~Karl Flöthner,~Dimitri Schaab,~Jonathan Ottnad,~Viktor Ratza,~Markus Ball,~Bernhard Ketzer\\
	Helmholtz-Institut für Strahlen- und Kernphysik\\
	Universität Bonn\\
	53115 Bonn - Germany
}

\begin{document}
	\maketitle
	
	\begin{abstract}
			Gas Electron Multipliers (GEM) are widely used as amplification stages in gaseous detectors exposed to high rates. 
		The GEM consists of a polyimide foil which is coated by two thin copper layers. 
		Charges collected into the holes or created during the amplification process may adhere to the polyimide part inside the holes.
		This is often accompanied by a change of the effective gain.
		The effect is commonly known as the~\enquote{charging-up effect}.
		
		This work presents a systematic investigation of the effect under well-controlled and monitored conditions for a single standard GEM foil in Ar/CO$_2$ (90/10) gas for varying rates of X-ray interactions in the detector and for different gains of the foil.
		In order to cover a wide range of different rates, we apply two different methods.
		The first one is based on a current measurement while the second one relies on the analysis of $^{55}$Fe spectra over time.
		Both methods give consistent results, showing an initial increase of the effective gain with time and an asymptotic saturation, which can be well described by a single-exponential function. 
		We find that the characteristic time constants extracted from our measurements scale inversely proportional to the rate of incoming electrons for a given GEM voltage. 
		Introducing characteristic quantities, which describe either the number of incoming electrons per hole or the total number of electrons in a hole per characteristic time, we find consistent numbers for measurements taken at the same GEM voltages. 
		For measurements taken at different GEM voltages, however, also the characteristic total number of electrons inside the hole, which is supposed to take into account the different effective gains, is found to be higher by a factor of about 3.5 for $U_\mathrm{GEM}=\SI{350}{\volt}$ ($n_\mathrm{tot}=8.8\times 10^8$) compared to \SI{400}{\volt} ($n_\mathrm{tot}=2.4\times 10^8$). 
		This hints at a residual dependence of the charging-up characteristics on the GEM voltage.
	\end{abstract}


\section{Introduction}
\label{Sec:Intro}
\begin{figure}
	\hfill
	\begin{minipage}{0.41\textwidth}
		\ifthenelse{\boolean{color}}
		{\includegraphics[width=\textwidth]{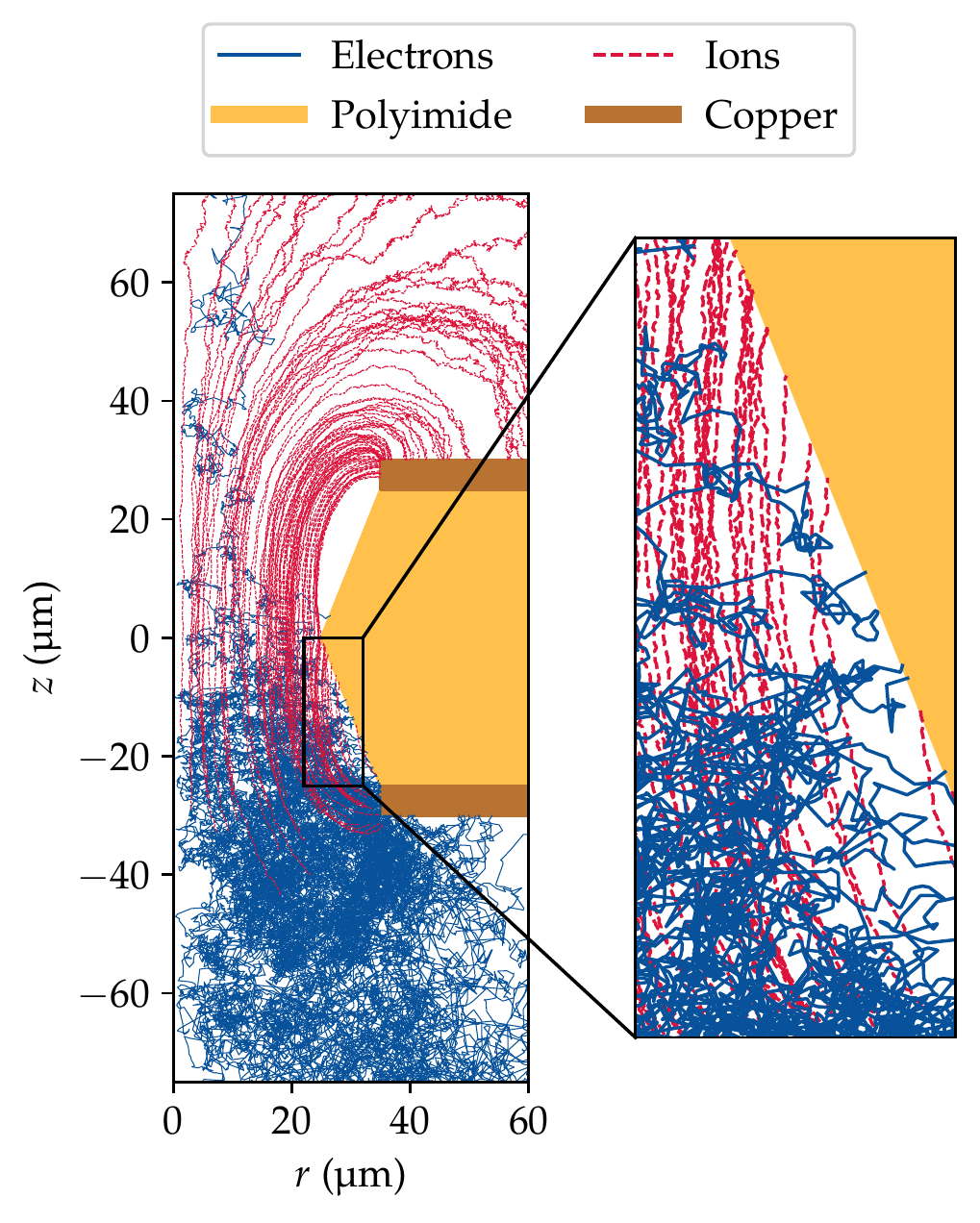}}
		{\includegraphics[width=\textwidth]{All_Particles_gray.pdf}}
	\end{minipage} 
	\hfill
	\begin{minipage}[c]{0.4\textwidth}
		\caption{\label{Fig:Avalanche_Simu}
			Simulated trajectories of electrons (solid lines) and ions (dashed lines) in a single GEM hole.
			The close-up view shows how particles end up on the polyimide.
			For this simulation, a single electron was placed \SI{100}{\micro\meter} above the GEM.
			A drift field of \SI{400}{\volt\per\centi\meter} lets the electron drift towards the GEM hole.
			The potential difference over the GEM was chosen to be~\SI{300}{\volt}.
			To extract electrons, a field of~\SI{2000}{\volt\per\centi\meter} was applied below the GEM.
			The gas was set to a mixture of argon and carbon-dioxide (90:10).\newline
			To simulate the movement of electrons and ions, the framework Garfield\texttt{++}~(a C\texttt{++} adaption of the Fortran-based framework Garfield~\cite{VEENHOF1998726}) was used.
			ANSYS\trademark{} was used to calculate the electrostatic environment. Color online.
			\vspace*{1cm}}
	\end{minipage}\hspace{\fill}
\end{figure}

Detectors based on the Gas Electron Multiplier (GEM)~\cite{SAULI1997531} are widely used in particle physics experiments that require high position resolution over large areas in high-rate environments~(e.g. COMPASS~\cite{KETZER2002142, ALTUNBAS2002177}, LHCb~\cite{BENCIVENNI2002493}, TOTEM~\cite{BAGLIESI2010134, 1462231}, JLab Hall A~\cite{GNANVO201577} as well as ALICE~\cite{KETZER2013237,LIPPMANN2016543} and CMS~\cite{ABBANEO2013383,CALABRIA20161042} after their upgrades).
The GEM consists of a \SI{50}{\micro\meter} thick polyimide foil which is coated on both sides with~\SI{5}{\micro\meter} thick copper layers.
In a photolithographic process, holes are etched into this foil in a hexagonal pattern.
Standard GEM foils have an inner diameter of approximately~\SI{50}{\micro\meter}, an outer diameter of approximately~\SI{70}{\micro\meter} and a pitch between two neighbouring holes of~\SI{140}{\micro\meter}.
If a suitable voltage is applied between both copper layers, strong non-uniform electric fields are created inside the holes (of the order of~\SI{50}{\kilo\volt\per\centi\meter}), sufficient for incoming electrons to start an avalanche of further ionizations.
During this multiplication process, electrons and ions may diffuse to the polyimide part of the GEM and be adsorbed there as shown in figure~\ref{Fig:Avalanche_Simu}.
Due to the high resistivity of the material, the charges remain there for a rather long time.
These new charges accumulate over time and dynamically change the electric field inside the hole.
This is known as the \enquote{charging-up effect}.
Many publications suggest that the charging-up effect is responsible for a change of the effective gain  which is time-dependent and reaches asymptotically a constant value (e.g. in measurements with GEMs~\cite{ALTUNBAS2002177, 4395404, 4179872, Mythra, Croci_Gain_Decrease}, in measurements with different micropattern gaseous detectors~\cite{Pitt_Measurement, Renous_Measurements}, as well as in simulations~\cite{Correia_Simulations, Correia_Simulations_2, Alfonsi_Simulations}).
A quantitative understanding of the various effects reported (which e.g. include both an increase~\cite{ALTUNBAS2002177} and a decrease~\cite{Croci_Gain_Decrease} of the effective gain), however, has not been reached in our opinion.
In addition, as we had to experience ourselves, there are many other external effects which may mimic a genuine charging-up effect.
Examples are changes in external conditions like temperature and pressure, initial instabilities of X-ray generators or time constants intrinsic to the high-voltage power supply system (see also section~\ref{Sec:Setup}).
Especially for applications that require the gain to remain very stable over time, e.g. in Time Projection Chambers for d$E$/d$x$ measurements~\cite{KETZER2013237, BERGER2017180} or in photon detectors~\cite{Meinschad_Photon_Detectors}, a quantitative understanding of the effect is indispensable.

This work presents a systematic investigation of the charging-up effect under well-controlled and monitored conditions.
By using two different methods, we are able to investigate the time-constant over a wide range of rates of X-ray interactions.
The first method makes use of a conventional X-ray tube where the amplified ionization currents are sufficiently large to be measured with a picoamperemeter.
For the second method, the peak position of the $K_\alpha$ line in an $^{55}$Fe spectrum is observed over time.
Both types of measurement were conducted using a single GEM as amplification stage~\cite{Benlloch_Measurement_Methods}.

	\section{Setup}
	\label{Sec:Setup}
\begin{figure}
		\centering
		\begin{tikzpicture}
		\draw[very thick] (1, 4.5) -- (1,4) node[pos=0.5, left]{Gas in} -- (0,4) -- (0,0) -- (8,0) -- (8,4) node[pos=0.5, below, rotate=90]{Aluminum Vessel} -- (7,4) -- (7,4.5) node[pos=0.5, right]{Gas out};
		\draw[thick, arrows=->] (1.25, 4.5) -- (1.25, 4);
		\draw[thick, arrows=->] (6.75, 4) -- (6.75, 4.5);
		\draw[very thick] (1.5, 4.5) -- (1.5,4) -- (2,4);
		\draw[very thick, color=copper] (2,4) -- (6,4) node[pos=0.5, below]{Kapton Window};
		\draw[very thick] (6,4) -- (6.5,4) -- (6.5,4.5);
		
		\draw[fill] (3.5, 4.2) -- (3.9, 4.2) -- (3.9, 5) -- (3.5, 5) -- (3.5, 4.2);
		\draw[fill] (4.5, 4.2) -- (4.1, 4.2) -- (4.1, 5) -- (4.5, 5) -- (4.5, 4.2) node[pos=0.5, right]{Collimator};
		
		\draw[fill, color=rad_yellow] (4, 5.6) -- (4.3, 5.1) -- (3.7, 5.1) -- cycle;
		\draw[line width=0.8] (4, 5.6) node[above]{X-ray tube or $^{55}$Fe source} -- (4.3, 5.1)  -- (3.7, 5.1) -- cycle;
		\draw[line width=0, fill] (4 + 0.5*0.145, 5.275 + 0.866025*0.145) -- (4 + 0.145, 5.275) arc (0:60:0.145) -- cycle;
		\draw[line width=0, fill] (4 + 0.5*0.145, 5.275 + 0.866025*0.145) -- (4, 5.275) -- (4 + 0.145, 5.275) -- cycle;
		\draw[line width=0, fill] (4 + 0.5*0.145, 5.275 - 0.866025*0.145) -- (4 - 0.5*0.145, 5.275 - 0.866025*0.145) arc (240:300:0.145) -- cycle;
		\draw[line width=0, fill] (4 + 0.5*0.145, 5.275 - 0.866025*0.145) -- (4, 5.275) -- (4 - 0.5*0.145, 5.275 - 0.866025*0.145) -- cycle;
		\draw[line width=0, fill] (4 - 0.5*0.145, 5.275 + 0.866025*0.145) -- (4 - 0.145, 5.275) arc (180:120:0.145) -- cycle;
		\draw[line width=0, fill] (4 - 0.5*0.145, 5.275 + 0.866025*0.145) -- (4, 5.275) -- (4 - 0.145, 5.275) -- cycle;
		\draw[fill, color=rad_yellow] (4, 5.275) circle [radius=0.035];
		\draw[fill, color=black](4, 5.275) circle [radius=0.02];
		
		\draw[line width = 2.5, color=copper] (2,3) -- (6,3) node[pos=1, right, text=black]{Drift Foil};
		\draw[line width = 1, color=gold](2,3.05) -- (6,3.05);
		\draw[line width = 1, color=gold](2,2.95) -- (6,2.95);
		
		\draw[line width = 2.5, color=copper] (2,1.2) -- (6,1.2) node[pos=1, right, text=black]{GEM Foil};
		\draw[line width = 1, color=gold](2,1.25) -- (6,1.25);
		\draw[line width = 1, color=gold](2,1.15) -- (6,1.15);
		

		\draw[line width = 2.5, color=gold](2,0.6) -- (6,0.6) node[pos=1, right, text=black]{Pad Plane};
		
		\draw[color=uni_grau, thick,decorate,decoration={brace,mirror}] (1.8, 2.95) -- (1.8, 1.25) node[midway, left, xshift=-2pt] {\SI{25.7}{\milli\meter}};
		\draw[color=uni_grau, thick,decorate,decoration={brace,mirror}] (1.8, 1.15) -- (1.8, 0.6) node[midway, left, xshift=-2pt] {\SI{2.25}{\milli\meter}};
		
		\end{tikzpicture}
		\caption{\label{Sketch_test_detector}Sketch of the used detector (not to scale).
			In total, the detector has a volume of approximately \SI{4}{\liter}.
			Immediately downstream of the gas outlet, a temperature and pressure sensor as well as a measurement device for the oxygen and water content are included in the gas line.}
	\end{figure}
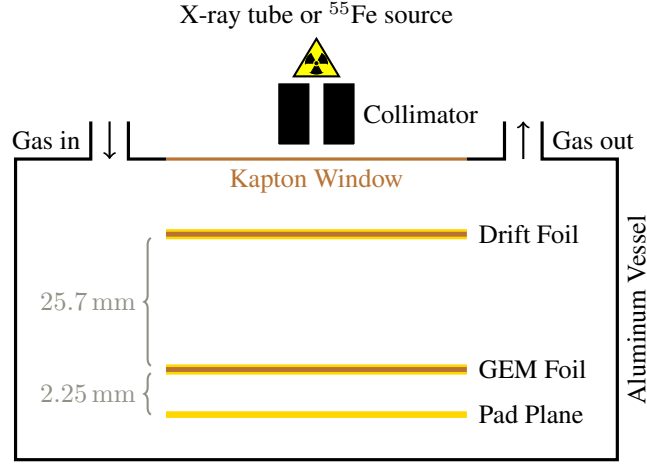

	In order to measure the charging-up effect, a dedicated detector was set up with a single GEM foil as amplification stage.
	A sketch of the used detector is shown in figure~\ref{Sketch_test_detector}.
	It consists of an aluminum vessel with a \SI{100}{\micro\meter} thick Kapton\trademark{} 400FN022\footnote{This type of foil was chosen because of its low water vapor permeability.} window on the top side to irradiate the detector with X-rays.
	The detector was constantly flushed with Ar/CO$_2$ (90/10) with a flow of \SI{3}{\liter\per\hour}.
	The total gas volume is approximately~\SI{4}{\liter}.
	It consists of a \SI{25.7}{\milli\meter} long drift volume, in which photons of an X-ray source can convert.
	The drift field~$E_\textrm{drift}$ is created by a \SI{50}{\micro\meter} thick polyimide foil, cladded on both sides with~\SI{5}{\micro\meter} copper\footnote{ The same base material as for GEMs, but without the hole pattern.}.  
	It was set to~\SI{400}{\volt\per\centi\meter} in all measurements.
	The used GEM foil with an active area of~$10\times\SI{10}{\square\centi\meter}$ was produced at CERN with the so-called \enquote{double-mask-technique}.
	With this technique, a double-conical shape of the holes is achieved where the diameter of the holes in the polyimide (called inner diameter) is slightly smaller than the diameter of the copper holes (called outer diameter).
	Since earlier measurements report a strong influence of the hole shape on the characteristics of the charging-up effect~\cite{Sauli:2016}, the outer and inner diameters of a few hundred holes were measured with an optical microscope.
	The results are: $d_{\textrm{inner}} = \SI{52.7\pm0.8}{\micro\meter}$, $d_{\textrm{outer}}^{\textrm{top}} = \SI{70.7 \pm 0.8}{\micro\meter}$ and~$d_{\textrm{outer}}^{\textrm{bot}} = \SI{72.8 \pm 0.9}{\micro\meter}$, where the uncertainties represent the RMS values of the measured distribution, and the superscripts \enquote{top} and \enquote{bot} refer to the top side (facing the drift electrode) and the bottom side (facing the readout electrode), respectively. 
	The voltage across the GEM was set to be either~\SI{350}{\volt} or~\SI{400}{\volt}.
	An induction field~$E_\textrm{ind}$ of~\SI{2000}{\volt\per\centi\meter} extracts the electrons from the GEM and guides them to a pad plane, segmented in 100 pads of $1\times\SI{1}{\square\centi\meter}$ size.
	The induced signals can then be read out for example with a picoamperemeter or can get amplified with a charge-sensitive amplifier, in order to analyze the pulse-height spectrum.
	The high voltage for each channel was supplied by a ISEG\trademark \emph{EHS 8060n} module.
	Each channel has a~\SI{1}{\giga\ohm} resistor to ground in order to set the ground reference for the system correctly, as the used high-voltage power supply can not sink currents.
	
	As an X-ray source, either a conventional X-ray tube (\emph{Mini-X} by Amptek\trademark) or an $^{55}$Fe source (with an activity of $\approx\SI{11.5}{\mega\becquerel}$) were used.
	The X-ray beam was collimated (see figure~\ref{Sketch_test_detector}), with different collimators being used for the measurements at different GEM voltages.
	We took great care to ensure that the collimator was oriented perpendicular to the GEM surface.
	In addition, we used collimators with a very small aspect ratio (diameter/length) to suppress the influence of the penumbra that arises around the central beam spot. In the region of the penumbra, the rate of X-rays and therefore also the rate of initial electrons is gradually decreasing, leading to a superposition of many different time constants for the charging-up effect, which would be impossible to disentangle. 
	A negligibly small penumbra is hence necessary for a proper investigation of the charging-up effect. 
	With the collimators used, the divergence of the X-ray beam and the penumbra are negligible, such that the irradiated area corresponds to the opening area of the collimator. 
	This was verified by illuminating Polaroid films with the collimated X-ray sources.
	
	With this setup, it is possible to investigate the charging-up effect especially with a focus on the characteristic time constant of the gain variations with respect to the rate of ionization electrons entering the GEM (determined by the rate of X-ray interactions in the conversion volume).
	Since the used $^{55}$Fe source creates a rather small rate of ionization electrons compared to the Mini-X, a huge bandwidth (from a few fA up to pA) can be covered.
	
	In order to assure that the measured gain variations are caused by the charging-up effect, several disruptive effects were carefully considered and ruled out.
	All measurements were conducted in a Faraday cage in order to reduce the influence of external noise sources on the signals.
	The voltage across the GEM was applied at least \SI{24}{\hour} before a measurement was conducted.
	Additionally, successive measurements were performed at different positions of the foil in order to ensure that the results were not biased due to partial or full charging-up of foils at a particular irradiation position. Indeed, performing two successive measurements at the same spot, we did not observe charging-up effects during the second measurement, proving that the measured change of the effective gain is a local effect. 
	As was shown in \cite{ALTUNBAS2002177}, the charging-down time is of the order of several hours, which would require very long waiting times for measurements at the same spot. 
	Our measurements confirm this behaviour (see also Sec.~\ref{Sec:Discussion}).
	Temperature, pressure, humidity and oxygen content of the gas were constantly monitored.
	All these environmental parameters were measured in the gas system, directly after the detector.
	Another important issue is a constant rate of X-ray interactions in the detector. Since the Mini-X needs some operational time until it delivers a constant rate, a shutter blocked the X-ray beam from entering the detector for all measurements with the Mini-X, until the rate of X-rays was stable (after approx.~\SI{60}{\minute}).
	
\section{Measurement Methods}
\subsection{Measurement method \rom{1} -- Current measurement}\label{Sec:Explanation_of_Measurement_Method_1}
The first measurement method makes use of a picoamperemeter (originally developed at TU München and further improved at Bonn University~\cite{Schaab_pAPoster}) which was connected to the pad plane of the detector.
In its most sensitive mode, the picoamperemeter has a digital resolution of \SI{0.5}{\pico\ampere} and an absolute accuracy of~\SI{2}{\pico\ampere}, given by a known temperature dependence and possible calibration uncertainties.
Over a time interval of approximately \SI{10}{\second}, it measures 128 values and sends out the average value as well as the standard deviation (which yields the uncertainty of the average value when divided by $\sqrt{128}$, assuming that the fluctuations are purely statistical).
A known residual temperature-dependent behaviour of the picoamperemeter did not affect single measurements, since the temperature variations were always smaller than \SI{0.5}{\degreeCelsius}.
For measurements performed at different times, possible shifts of the order of a few pA due to different ambient temperatures are taken into account in the systematic uncertainty.

In order to create measurable currents, the Mini-X was used as irradiation source.
Here, the GEM voltage was either set to~\SI{350}{\volt} or~\SI{400}{\volt}.
The effective gain~$G_\textrm{eff}$ of the GEM is defined as the ratio between the 
current measured on the readout anode~$I_\textrm{readout}(t)$ and the ionization current~$I_\textrm{ionization}$ induced by X-ray interactions in the drift volume:
\begin{align}\label{Eq:Effective_Gain}
G_\textrm{eff}\! \left( t \right) = \frac{I_\textrm{readout}\! \left( t \right) }{I_\textrm{ionization}} \; .
\end{align}
The ionization current can be measured on the top side of the GEM with the GEM voltage and the induction field set to zero.
For the charging-up measurements, however, the ionization currents were too small to be measured directly.
Instead, we measured the effective gain separately using higher ionization currents by increasing the current of the Mini-X significantly.

The range of possible X-ray interaction rates is mainly limited by the accuracy of the picoamperemeter.
If the current of the Mini-X is too small, the readout current~$I_\textrm{readout}$ is also small and can not be measured reliably by the picoamperemeter.
If the Mini-X current is too high, the time-constant of the charging-up effect is too small to be measured since the used picoamperemeter measures only one value each ten seconds.
Hence, in the next section, a different method is presented that makes use of smaller X-ray interaction rates.

\subsection{Measurement method \rom{2} -- Pulse-height spectrum analysis}\label{Sec:Explanation_of_Measurement_Method_2}
The second measurement method is based on measuring the pulse-height spectrum from an $^{55}$Fe source.
Since the effective gain of a single GEM at~\SI{350}{\volt} in Ar/CO$_2$ (90/10) is not sufficient to observe a clean spectrum, this measurement method could only be used at a GEM voltage of~\SI{400}{\volt}.
A single-channel charge-sensitive pre-amplifier~(\emph{Ortec 142}~\cite{Ortec142}) was connected to the pad plane, with the signals from the four innermost pads summed up (the area of the four pads is \SI{4}{\square\centi\meter}).
Afterwards, the signal was fed to the main amplifier~(\emph{Ortec 671}~\cite{Ortec671}) where it was shaped (shaping time \SI{3}{\micro\second}) and amplified (course gain 300, fine gain 1.5).
The shaped and amplified signal was then fed into a multi-channel analyzer (\emph{MCA-8000A}~\cite{MCA8000A}) which was connected to a computer.
The spectrum was then analyzed in order to extract the position of the $K_\alpha$ peak; more details will be given in section~\ref{Sec:Results_Second_Method}.

If the charging-up effect influences the gain of the GEM, the peak position of the $K_\alpha$ line  will vary over time.
Therefore, many spectra were recorded over a time period of several hours.
Each spectrum had a measurement time of either one or five minutes.
Since the rate of initial ionization electrons is small, the complete measurement took approximately~\SI{12}{\hour}.
On this timescale, temperature and pressure variations may influence the gain behaviour significantly.
Hence, the peak-position of the $K_\alpha$ line has to be corrected for these effects.
In order to do this, many spectra were recorded after the gain had saturated.
Making use of recorded variations of pressure~$p$ and temperature~$T$ in the detector, a correlation between the peak position of the $K_\alpha$ line and $T/p$ was derived.
A linear fit was applied to these data points and was then be used to correct for temperature and pressure variations.
\section{Results}\label{Sec:Results}
\subsection{Measurement method~\rom{1}}
\begin{figure}\centering
	\begin{minipage}{\textwidth}
		\hspace{\fill}
		\begin{minipage}{0.6\textwidth}
			\ifthenelse{\boolean{color}}
			{\includegraphics[width=\textwidth]{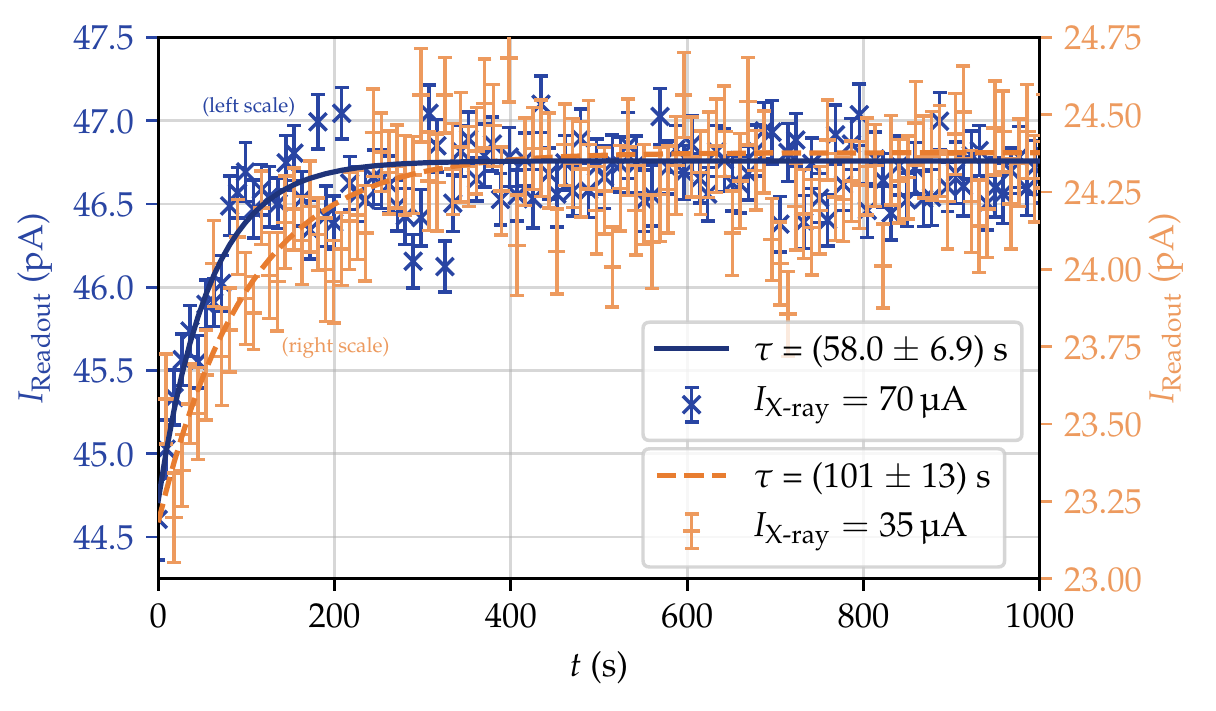}}
			{\includegraphics[width=\textwidth]{400V_Comparison_Paper_gray.pdf}}
		\end{minipage}\hfill
		\begin{minipage}[c]{0.38\textwidth}
			\caption{\label{Fig:Result_400V}
				Measured currents on the pad plane at a GEM voltage of~\SI{400}{\volt}.
				The darker data points (marked with a \enquote{x}) were taken while the Mini-X operated at~\SI{10}{\kilo\volt} and~\SI{70}{\micro\ampere} and they refer to the left axis.
				For the brighter data points (marked with a \enquote{+}) which refer to the right axis, the current of the X-ray tube was changed to~\SI{35}{\micro\ampere}.
				The error bars of the measured values refer to the statistical measurement uncertainty of the picoamperemeter (see section~\ref{Sec:Explanation_of_Measurement_Method_1}).
				Both measurements were conducted one after the other on different spots on the same GEM.
				The gas temperature was almost stable at \SI{19}{\degreeCelsius} as well as the pressure at \SI{1028}{\hecto\pascal}.
				The oxygen content was \SI{31}{\ppm} while the water content was \SI{9}{\ppmv}. Color online.}
		\end{minipage}\hspace{\fill}
	\end{minipage}
	\begin{minipage}{\textwidth}
		\hfill
		\begin{minipage}{0.6\textwidth}
			\ifthenelse{\boolean{color}}
			{\includegraphics[width=\textwidth]{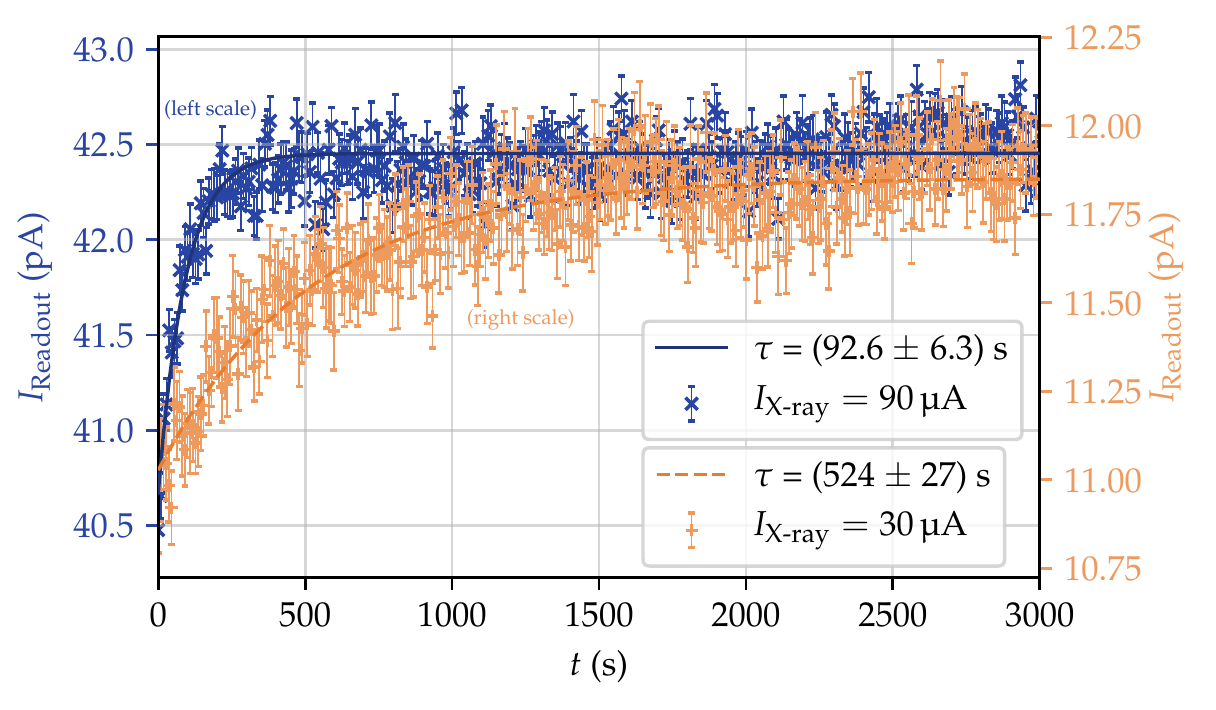}}
			{\includegraphics[width=\textwidth]{350V_Comparison_Paper_gray.pdf}}
		\end{minipage}\hfill
		\begin{minipage}[c]{0.38\textwidth}
			\caption{\label{Fig:Result_350V}
				Measured currents on the pad plane at a GEM voltage of~\SI{350}{\volt}.
				The darker data points (marked with a \enquote{x}) were taken while the Mini-X operated at~\SI{10}{\kilo\volt} and~\SI{90}{\micro\ampere} and they refer to the left axis.
				For the brighter data points (marked with a \enquote{+}) which refer to the right axis, the current of the X-ray tube was changed to~\SI{30}{\micro\ampere}.
				The error bars of the measured values refer to the statistical measurement uncertainty of the picoamperemeter (see section~\ref{Sec:Explanation_of_Measurement_Method_1}).
				Both measurements were conducted one after the other on different spots on the same GEM.
				The gas temperature was almost stable at \SI{22}{\degreeCelsius} as well as the pressure at \SI{1022}{\hecto\pascal}.
				The oxygen content was \SI{22}{\ppm} while the water content was \SI{27}{\ppmv}. Color online.}
		\end{minipage}
		\hspace{\fill}
	\end{minipage}
\end{figure}

The results from measurement method~\rom{1} are depicted in figures~\ref{Fig:Result_400V} and~\ref{Fig:Result_350V}, which show the currents measured at the readout anode as a function of time for different currents of the X-ray tube and for two different GEM voltages (\SI{400}{\volt} in figure~\ref{Fig:Result_400V} and \SI{350}{\volt} in figure~\ref{Fig:Result_350V}).
All measured currents exhibit an initial increase with time and a saturation after longer irradiation times.
To each data set, a single exponential function of the form
\begin{align}\label{Eq:Exponential_Gain_Increase}
I_{\textrm{readout}} \left( t \right) = I_{\textrm{sat}}  \left[1-\frac{\Delta I}{I_\textrm{sat}} \exp \left( - t / \tau \right)\right]
\end{align}
was fitted.
$I_{\textrm{readout}}$ denotes the measured current on the pad plane, $I_{\textrm{sat}}$ the saturation current, $\Delta I$ the increase of current during the charging-up process, and $\tau$ the time constant of the charging-up effect.
A summary of key quantities during the measurements and fit results are given in table~\ref{Tab:Measurement_Method_One_Summary}.

On first sight, it can be seen from figures~\ref{Fig:Result_400V} and~\ref{Fig:Result_350V} that an increased rate of ionization electrons (due to a higher current of the X-ray tube~$I_\textrm{X-ray}$) leads to a faster gain increase.
Since the effective gain is different for different GEM voltages, the two measurements will first be analyzed separately.

For $U_\textrm{GEM} = \SI{400}{\volt}$, two measurements were conducted where the rate of ionization electrons differed by a factor of approximately two (since the current of the X-ray tube~$I_\textrm{X-ray}$ was decreased from~\SI{70}{\micro\ampere} to \SI{35}{\micro\ampere}).
To be more precise, the ratio of the saturation currents~$I_\textrm{sat}$ can be used as a measure for the ratio of the ionization currents, since~$I_\textrm{ionization} = I_\textrm{sat}/G_{\textrm{eff}}^{\textrm{sat}}$, where $G_{\textrm{eff}}^{\textrm{sat}}$ denotes the gain after full charging-up (see Eq.~\ref{Eq:Effective_Gain}):
\begin{align}\label{Eq:Quotient_Charge_Up_Sat_Currents}
\frac{I_\textrm{sat}^{\SI{35}{\micro\ampere}}}{I_\textrm{sat}^{\SI{70}{\micro\ampere}}} 
=
\frac{I_\textrm{ionization}^{\SI{35}{\micro\ampere}} \cdot G_\textrm{eff}^{\textrm{sat}}}
{I_\textrm{ionization}^{\SI{70}{\micro\ampere}} \cdot G_\textrm{eff}^{\textrm{sat}}}
= 
\frac{I_\textrm{ionization}^{\SI{35}{\micro\ampere}}}{I_\textrm{ionization}^{\SI{70}{\micro\ampere}}}
= 
\num{0.52 \pm 0.06}~\textrm{(syst.)} \; .
\end{align}
A possible change of the saturated effective gain (e.g. by a position dependence or by changing environmental parameters like temperature or pressure) is included in the systematic uncertainty, which is estimated conservatively by dividing the maximum possible value ($I_\textrm{sat}^{\SI{35}{\micro\ampere}} + \SI{2}{\pico\ampere}$) by the minimum possible value ($I_\textrm{sat}^{\SI{70}{\micro\ampere}} - \SI{2}{\pico\ampere}$) and vice versa.
A similar estimation of the systematic uncertainty is also done for other quantities presented in this work.
The statistical uncertainty is much smaller than the quoted systematic uncertainty and hence neglected.

The ratio of time constants for the two measurements at different X-ray currents is
\begin{align}
\frac{\tau^{\SI{70}{\micro\ampere}}}{\tau^{\SI{35}{\micro\ampere}}} = \num{0.54 \pm 0.09}~\textrm{(stat.)} \; ,
\end{align}
where the statistical uncertainty was calculated with Gaussian error propagation, taking into account the fit uncertainties.
Both values agree within uncertainties, suggesting that for a fixed GEM voltage, the charging-up time constant scales inversely proportional to the ionization current, or the rate of X-ray interactions, as might have been naively expected.

For $U_\textrm{GEM} = \SI{350}{\volt}$, the same calculation can be done.
Note that the current of the X-ray tube differs by a factor of three for these measurements.
The results are
\begin{align}
\frac{I_\textrm{sat}^{\SI{30}{\micro\ampere}}}{I_\textrm{sat}^{\SI{90}{\micro\ampere}}} = \num{0.28 \pm 0.06}~\textrm{(syst.)} \qquad \textrm{and} \qquad \frac{\tau^{\SI{90}{\micro\ampere}}}{\tau^{\SI{30}{\micro\ampere}}} = \num{0.177 \pm 0.015}~\textrm{(stat.)} \; .
\end{align}
Although the difference is slightly larger than one sigma, the agreement can still be considered fair.

\begin{table}
	\centering
	\caption{\label{Tab:Measurement_Method_One_Summary}Comparison of all measured values, obtained with the measurement method~\rom{1}.
		The uncertainty values  for $\tau$ and $I_{\textrm{sat}}$ are given by the fit uncertainty.
		For the calculation of the uncertainty of $n_\mathrm{inc}$ and $n_\mathrm{tot}$, the systematic uncertainties for $I_{\textrm{sat}}$, $G_{\textrm{eff}}^{\textrm{sat}}$ and $A_{\textrm{irr}}$  are taken into account, as explained in the text.
		or these quantities, the first uncertainty is statistical, while the second is systematic.
	}
	\begin{tabular}{@{}
			c
			S[table-format=3.0]@{\,\( \pm \)\,}S[table-format=2.0]
			c
			S[table-format=3.1]@{\,\( \pm \)\,}S[table-format=2.1]
			S[table-format=2.3]@{\,\( \pm \)\,}S[table-format=1.3]
			c
			@{\hspace{1em}(}S[table-format=2.2]@{\,\( \pm \)\,}S[table-format=1.2]@{\;}
			l@{)$\times10^8$}
			@{}}
		\toprule
		$U_{\textrm{GEM}}$ & \multicolumn{2}{c}{$G_{\textrm{eff}}^{\textrm{sat}}$} & $I_{\textrm{X-ray}}$ & \multicolumn{2}{c}{$\tau$} & \multicolumn{2}{c}{$I_{\textrm{sat}}$} & \multicolumn{1}{c}{$n_{\textrm{inc}}$} & \multicolumn{3}{c}{$n_\textrm{tot}$}\\ 
		(V) & \multicolumn{2}{c}{} & (\si{\micro\ampere}) &  \multicolumn{2}{c}{(s)} & \multicolumn{2}{c}{(\si{\pico\ampere})} & \multicolumn{1}{c}{(\si{\e\per\hole})} & \multicolumn{3}{c}{(\si{\e\per\hole})}\\
		\midrule
		400 & 150 & 10 & 35 & 107   & 13   &  24.379 & 0.007 & $\left( 5.3 \pm 0.6 \;^{+1.4}_{-1.2}\right)\!\times\!10^5$ & 2.47 & 0.29 & $^{+0.5}_{-0.4}$ \\[0.5ex]
		400 & 150 & 10 & 70 &  58.0 & 06.9 &  46.758 & 0.008 & $\left( 5.5 \pm 0.6\;^{+1.3}_{-1.0}\right)\!\times\!10^5$ & 2.56 & 0.32 & $^{+0.4}_{-0.31}$ \\ \midrule
		350 & 45 & 5 & 30 & 524   & 27   &  11.852 & 0.008 & $\left( 8.3 \pm 0.4 \;^{+4.0}_{-2.8}\right)\!\times\!10^6$ & 10.6 & 0.6 & $^{+3.4}_{-2.9}$ \\[0.5ex]
		350 & 45 & 5 & 90 &  92.6 & 06.3 &  42.452 & 0.007 & $\left( 5.2 \pm 0.4\;^{+1.8}_{-1.3}\right)\!\times\!10^6$ &  6.9 & 0.6 & $^{+1.4}_{-1.1}$  \\ \bottomrule
	\end{tabular}
\end{table}

In order to compare measurements performed at different ionization rates (X-ray currents), a new quantity is introduced following \cite{Pitt_Measurement, Correia_Simulations_2} to describe the characteristics of the charging-up process in terms of charges rather than time: the number of incoming electrons per hole~$n_{\textrm{inc}}$, which is the number of ionization electrons that arrive at a given hole during the time span of one time constant~$\tau$.
It can be calculated from the measured quantities by
\begin{align}\label{Eq:n_eph_minix}
n_{\textrm{inc}} = \frac{I_{\textrm{sat}} \cdot \tau}{G_{\textrm{eff}}^{\textrm{sat}}\cdot A_{\textrm{irr}}\cdot \rho_{\textrm{hole}}\cdot e} \; ,
\end{align}
where $A_{\textrm{irr}}$ denotes the irradiated area, $\rho_{\textrm{hole}}$ the areal hole density of a GEM foil and $e$ the elementary charge.
Since $I_\mathrm{sat}$ scales linearly with the ionization current (or rate of X-ray interactions), $n_\mathrm{inc}$ is expected to be independent of the rate, if $\tau$ indeed scales inversely proportional to the rate.
The hole density~$\rho_{\textrm{hole}}$ of a standard GEM foil can be calculated by geometric considerations to be~\SI[per-mode=reciprocal]{58.91}{\per\milli\meter\squared}.
As explained in Sec.~\ref{Sec:Setup}, the irradiated area~$A_{\textrm{irr}}$ corresponds to the opening area of the collimator.
For the measurements at~$U_{\textrm{GEM}}$~=~\SI{400}{\volt}, a collimator with an opening radius~$r_{\textrm{coll}}^{\textrm{400V}}$ of $\SI{1.05 \pm 0.05}{\milli\meter}$ (length of the collimator \SI{32}{\milli\meter}) was used, while a collimator with radius~$r_{\textrm{coll}}^{\textrm{350V}}~=~\SI{0.75 \pm 0.05}{\milli\meter}$ (length \SI{12}{\milli\meter}) was used for the measurements at~$U_{\textrm{GEM}}$~=~\SI{350}{\volt}.

The effective gain was determined by measuring the ionization current on the top side of the GEM, while the voltage across the GEM was set to~\SI{0}{\volt} and the induction field to~\SI{0}{\volt\per\centi\meter}.
With the ionization current and the measured saturation current, the effective gain can be calculated, see equation~\ref{Eq:Effective_Gain}.
The resulting values are $G_{\textrm{eff}}^{\textrm{sat}} = \num{150 +- 10}$ at~$U_{\textrm{GEM}}$~=~\SI{400}{\volt} and $G_{\textrm{eff}}^{\textrm{sat}} = \num{45 +- 5}$ at~$U_{\textrm{GEM}}$~=~\SI{350}{\volt}, respectively.
The uncertainties are purely of systematic origin (mainly caused by the~\SI{2}{\pico\ampere} accuracy of the picoamperemeter), the statistical uncertainties due to the variation of the measured current values are negligible.

With this information, the number of incoming electrons per hole $n_{\textrm{inc}}$ can be calculated for each measurement.
The values are given in table~\ref{Tab:Measurement_Method_One_Summary}.
It can be seen that the values for $U_{\textrm{GEM}}~=~\SI{400}{\volt}$ as well as the ones for~$U_{\textrm{GEM}}~=~\SI{350}{\volt}$ agree within uncertainties.
For the two measurement series at different GEM voltages, \SI{400}{\volt} and~\SI{350}{\volt}, however, they differ by about one order of magnitude.
This indicates that the time constant of the charging-up effect depends on the total number of electrons rather than the initial number of electrons in a hole, as also discussed for thick GEMs in~\cite{Pitt_Measurement, Correia_Simulations_2}. 
We therefore define the total number of electrons in the GEM hole in a time span $\tau$ by
\begin{align}\label{Eq:n_tot_minix}
n_{\textrm{tot}} = n_\mathrm{inc} \frac{G_\mathrm{eff}^{\textrm{sat}}}{\varepsilon_\mathrm{ex}} \; ,
\end{align}
where $\varepsilon_\mathrm{ex}$ is the extraction efficiency of electrons, defined as the fraction of electrons inside a GEM hole, which is extracted into the induction gap~\cite{Killenberg}.
The extraction efficiency was extrapolated from measurements presented in~\cite{Ottnad_MPGD2017} to be \num{0.32 +- 0.01} for $U_{\textrm{GEM}}~=~\SI{400}{\volt}$ and \num{0.35 +- 0.01} for $U_{\textrm{GEM}}~=~\SI{350}{\volt}$.
The total number of electrons in the hole is also listed in Table~\ref{Tab:Measurement_Method_One_Summary}. 
One can see, that there is still a discrepancy by a factor of~\numrange{2.7}{4.3} (see table~\ref{Tab:Measurement_Method_One_Summary}).
This can be taken as a hint that there is a residual dependence of the charging-up effect on the GEM voltage, in addition to the total number of electrons.
This is further discussed in section~\ref{Sec:Discussion}.

\subsection{Measurement method~\rom{2}}\label{Sec:Results_Second_Method}
\begin{figure}
	\centering
	\hfill
	\begin{minipage}{0.45\textwidth}
		\ifthenelse{\boolean{color}}
		{\includegraphics[width=\textwidth]{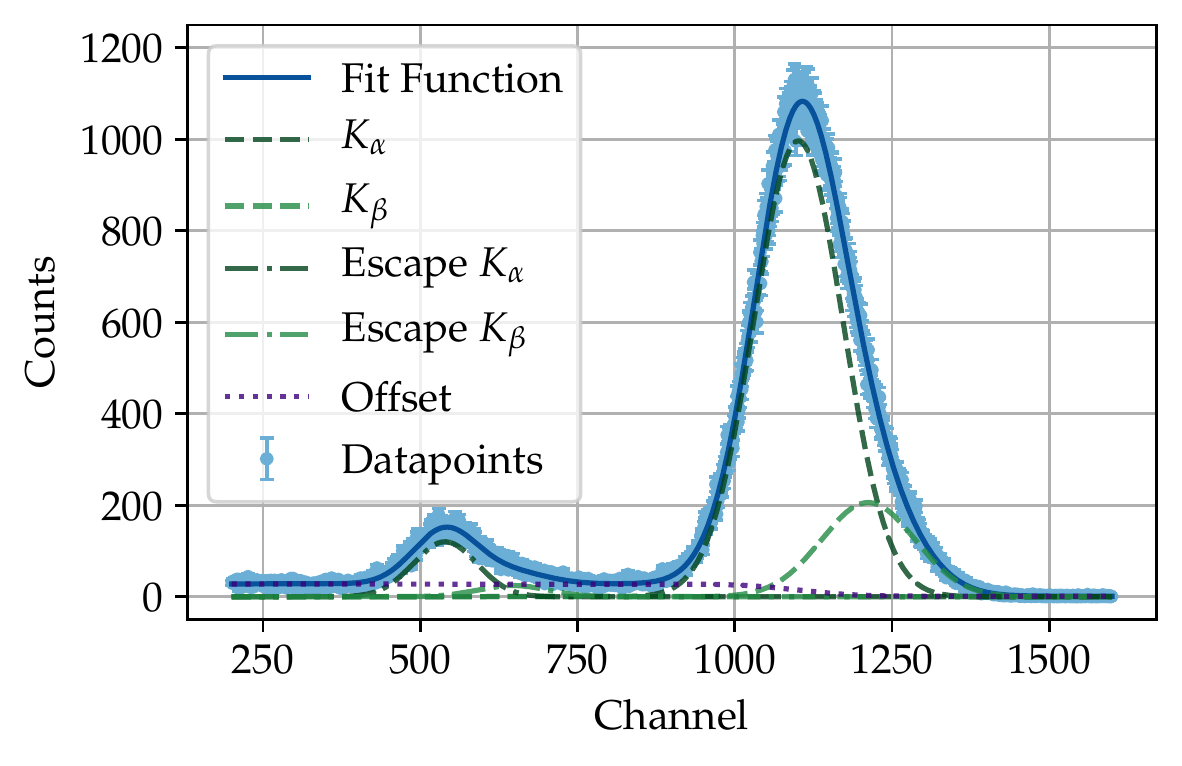}}
		{\includegraphics[width=\textwidth]{Plot_2019_06_11_600s_Spectrum_gray.pdf}}
		\captionsetup{format=hang, indention=0.15cm}
		\subcaption{\label{Fig:Fe55_Measurement_Spectrum_1}
			Spectrum taken after measurement (a).}
	\end{minipage}
	\hfill
	\begin{minipage}{0.45\textwidth}
		\ifthenelse{\boolean{color}}
		{\includegraphics[width=\textwidth]{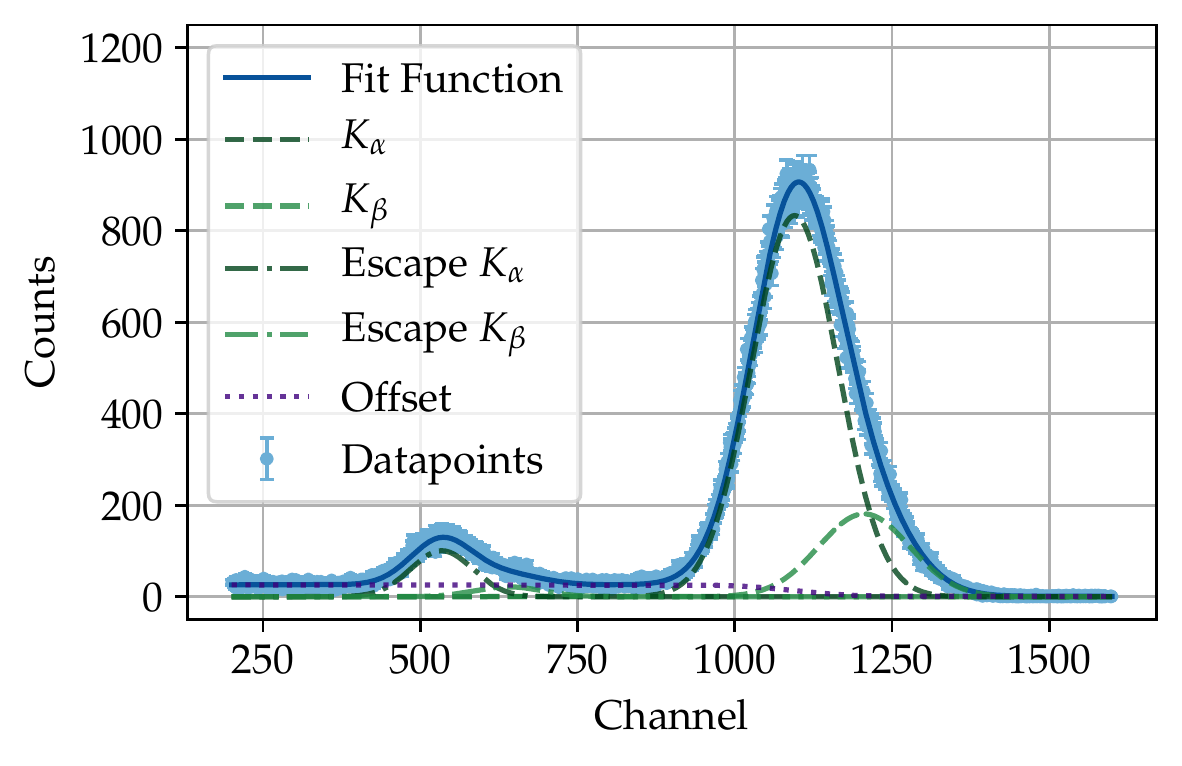}}
		{\includegraphics[width=\textwidth]{2019_09_20_600s_Spectrum_gray.pdf}}
		\captionsetup{format=hang, indention=0.15cm}
		\subcaption{\label{Fig:Fe55_Measurement_Spectrum_2}
			Spectrum taken after measurement (b).}
	\end{minipage}
	\hfill
	\caption{Two spectra --- each with a measurement time of~\SI{600}{\second} --- were taken after their respective charging-up measurement in order to calculate the ionization current.
		A physics-motivated fit-model was applied.
		The uncertainty for each data point is assumed to be~$\sqrt{N}$ where $N$ is the number of counts in this channel.
		Color online.
		\label{Fig:Fe55_Measurement_spectra}}
\end{figure}

\begin{figure}[!htb]
	\centering
	\hfill
	\begin{minipage}{0.45\textwidth}
		\ifthenelse{\boolean{color}}
		{\includegraphics[width=\textwidth]{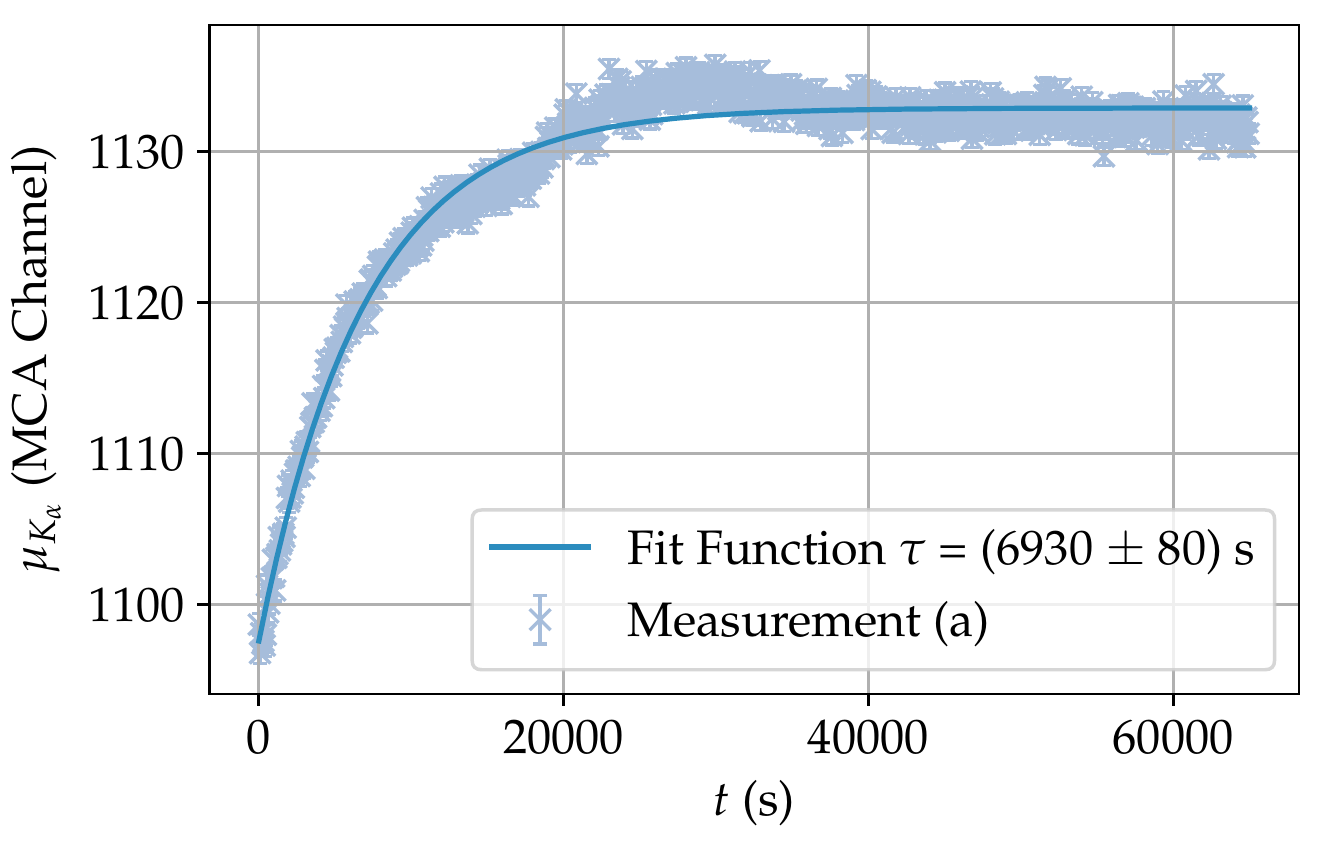}}
		{\includegraphics[width=\textwidth]{Charge_Up_Curve_2019_06_11_bw.pdf}}
		\captionsetup{format=hang}
		\subcaption{\label{Fig:Fe55_Measurement_1}First measurement with the $^{55}$Fe source.
			The recording time for each spectrum was set to~\SI{1}{\minute}.
			The reduced chi-square is $\chi^2_\textrm{red}~=~3.57$.\\
			During the measurement, the oxygen content was almost constant at \SI{30}{\ppm} while the water content decreased from~\SIrange{21}{19.5}{\ppmv}.\\
			The pressure increased from \SIrange{1014}{1018}{\hecto\pascal} and the temperature decreased from~\SIrange{19.6}{19.2}{\degreeCelsius}.}
	\end{minipage}
	\hfill
	\begin{minipage}{0.45\textwidth}
		\ifthenelse{\boolean{color}}
		{\includegraphics[width=\textwidth]{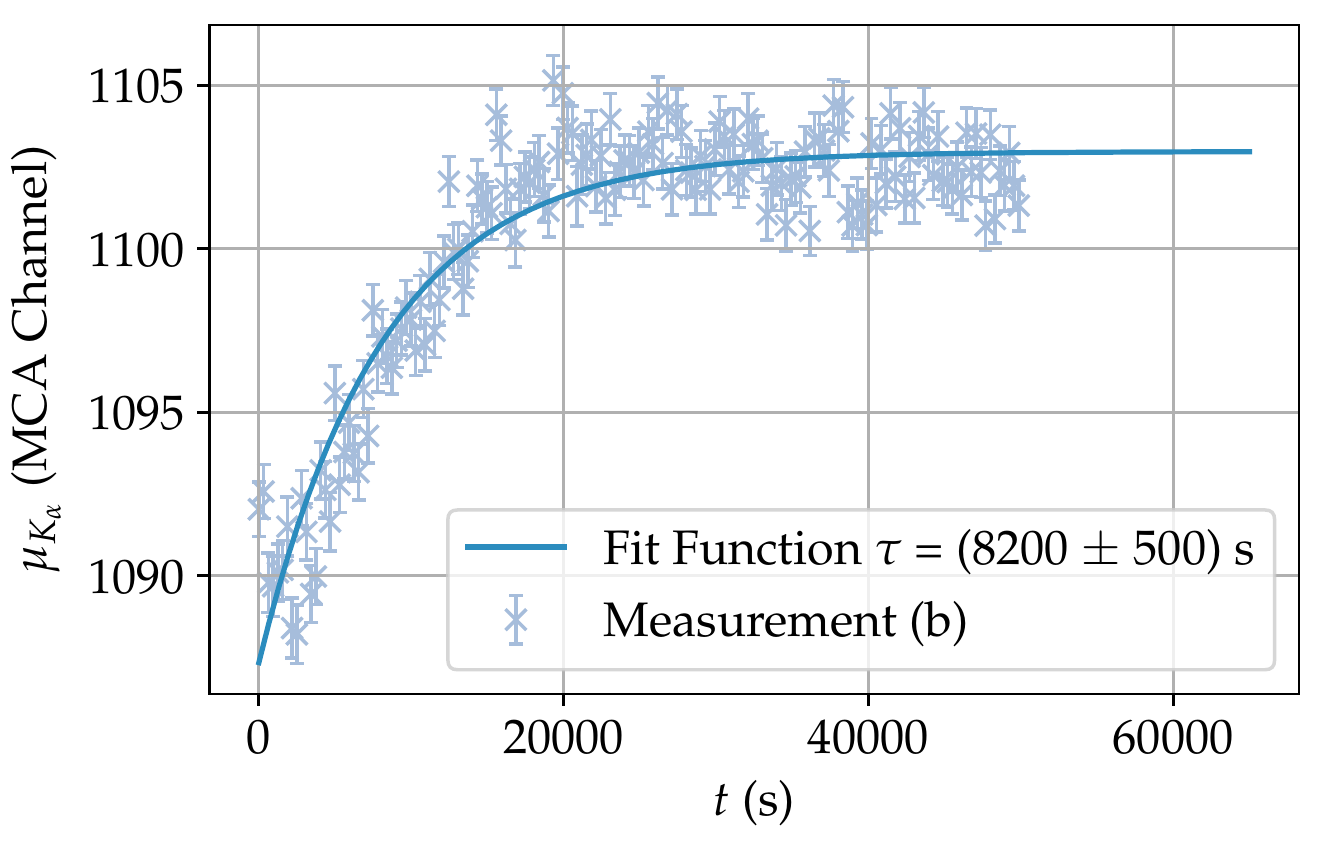}}
		{\includegraphics[width=\textwidth]{Charge_Up_Curve_2019_09_20_bw.pdf}}
		\captionsetup{format=hang}
		\subcaption{\label{Fig:Fe55_Measurement_2}Second measurement with the $^{55}$Fe source.
			The recording time for each spectrum was set to~\SI{5}{\minute}.
			The reduced chi-square is $\chi^2_\textrm{red}~=~2.93$.\\
			During the measurement, the oxygen content was almost constant at \SI{12.5}{\ppm} while the water content decreased from~\SIrange{61}{55}{\ppmv}.\\
			The pressure decreased from \SIrange{1034}{1028}{\hecto\pascal} and the temperature decreased from~\SIrange{22.4}{21.5}{\degreeCelsius}.
		}
	\end{minipage}
	\hfill
	\caption{Temperature and pressure-corrected peak positions of the $K_\alpha$ line of an $^{55}$Fe source as a function of time. Color online.
		\label{Fig:Fe55_Measurement}}
\end{figure}

Two measurements by method~\rom{2} were conducted which will be refered to as (a) and (b).
As explained in section~\ref{Sec:Explanation_of_Measurement_Method_2}, the recorded spectra were analyzed.
For this, a physics-based fit model was applied as shown for example in figure~\ref{Fig:Fe55_Measurement_spectra}.
It uses a superposition of four Gaussian-functions ($K_\alpha$, $K_\beta$ and their respective escape peaks) and in addition an error-function to describe the incomplete collection of charges.
The peak centers were fixed relative to each other, as well as the widths of the Gaussian peaks.
This can be done, since the energy of the lines and the energy dependence of the energy resolution\footnote{The relative energy resolution of a gaseous detector decreases proportional to 1/$\sqrt{E}$, where $E$ denotes the energy of the incident particle.} are known.
From the spectra, the centers $\mu_{K_\alpha}$ of the~$K_\alpha$ line can be extracted.
Afterwards, the data points were corrected for changes in temperature and pressure (see section~\ref{Sec:Explanation_of_Measurement_Method_2}).
For measurement (a) (depicted in figure~\ref{Fig:Fe55_Measurement_1}) the relation 
\begin{align}
\mu_{K_\alpha}^{(a)}\!\left( T/p \right) =  \left(\num{11761 +- 11}\right) \cdot T/p - \left(\num{2270 +- 30}\right)
\end{align}
was found, while the relation for measurement (b) (depicted in figure~\ref{Fig:Fe55_Measurement_2}) was found to be 
\begin{align}
\mu_{K_\alpha}^{(b)}\!\left( T/p \right) =  \left(\num{17001 +- 73}\right) \cdot T/p - \left(\num{3818 +- 21}\right) \; .
\end{align}

$T/p$ is given in units of \si{\kelvin\per\hecto\pascal} and the result is in units of MCA channels.
The $T/p$-corrected results are depicted in figure~\ref{Fig:Fe55_Measurement}, where the measurement time for one spectrum was set to one minute for figure~\ref{Fig:Fe55_Measurement_1} and to five minutes for figure~\ref{Fig:Fe55_Measurement_2}.
Due to a small change of the settings of the amplifier, the curves do not saturate at the same value.

Also with this method we observe a clear increase of the effective gain with time that can be fitted with a single exponential function in accordance with equation~\ref{Eq:Exponential_Gain_Increase}.
Since the rate of initial ionization electrons is smaller compared to the values from measurement method \rom{1}, the time constant is much longer.
Furthermore, the second measurement (figure~\ref{Fig:Fe55_Measurement_2}) was conducted approximately~100$\;$days after the first measurement (figure~\ref{Fig:Fe55_Measurement_1}).
Since the activity of the $^{55}$Fe source decreased over time, it also reasonable that a longer time constant is determined.

In order to compare the results from both measurement methods, the number of initial electrons per hole~$n_{\textrm{inc}}$ can be calculated for method~\rom{2}, too.
Here, the formula
\begin{align}\label{Eq:Calculation_e_per_hole_FE}
n_{\textrm{inc}} = \frac{I_{\textrm{ionization}} \cdot \tau}{A_{\textrm{irr}}\cdot \rho_{\textrm{hole}}\cdot e}
\end{align}
holds, with the same quantities as defined for equation~\ref{Eq:n_eph_minix}.
For both measurements, the same collimator was used which has an opening radius of~\SI{2.50 \pm 0.05}{\milli\meter}.
$I_{\textrm{ionization}}$ is the ionization current, created by the conversion of X-ray photons that are emitted by the $^{55}$Fe source.
This quantity can be derived by analyzing the pulse-height spectrum.
In figure~\ref{Fig:Fe55_Measurement_spectra}, two spectra are shown, where the first spectrum in figure~\ref{Fig:Fe55_Measurement_Spectrum_1} was recorded after the first measurement (figure~\ref{Fig:Fe55_Measurement_1}) and the second spectrum in figure~\ref{Fig:Fe55_Measurement_Spectrum_2} after the second measurement (figure~\ref{Fig:Fe55_Measurement_2}) took place.
The ionization current~$I_{\textrm{ionization}}$ can now be derived with the formula
\begin{align}\label{Eq:Determination_ionization_current}
I_{\textrm{ionization}} = \frac{A_{K_\alpha} E_{K_\alpha}  + A_{K_\beta} E_{K_\beta} + A_{K_\alpha^\textrm{esc}} E_{K_\alpha^\textrm{esc}} + A_{K_\beta^\textrm{esc}} E_{K_\beta^\textrm{esc}}}  {t_{\textrm{meas}} w_{\textrm{gas}}} \cdot e \; ,
\end{align}
where $A$ denotes the area under the respective peak and $E$ the energy of the line (in units of eV).
For each spectrum, the measurement time~$t_{\textrm{meas}}$ was set to~\SI{600}{\second}.
The mean energy required to create an electron-ion pair in an Ar/CO$_2$ (90/10) mixture~$w_{\textrm{gas}}~=~\SI{26.7}{\electronvolt}$~\cite{Sauli:117989}.

If all quantities are inserted into equation~\ref{Eq:Calculation_e_per_hole_FE}, the values
\begin{align}\label{Eq:Results_Measurement_Method_2}
n_{\textrm{inc}}^{\textrm{(a)}} = \SI{5.06 \pm 0.21 e5}{\e\per\hole} \qquad \textrm{and} \qquad  n_{\textrm{inc}}^{\textrm{(b)}} = \SI{4.9 \pm 0.4 e5}{\e\per\hole}
\end{align}
can be calculated.
They are in good agreement within their uncertainties.

\begin{table}
	\centering
	\caption{\label{Tab:Measurement_Method_Two_Summary}
		Comparison of all measured values, obtained with the measurement method~\rom{2}.
		The uncertainty values for $I_{\textrm{ionization}}$ and $\tau$ are given by the fit uncertainties.
		For the calculation of the uncertainty of $n_\mathrm{inc}$ and $n_\mathrm{tot}$, the systematic uncertainties for $G_{\textrm{eff}}^{\textrm{sat}}$ and $A_{\textrm{irr}}$  are taken into account, as explained in the text.
		For these quantities, the first uncertainty is statistical, while the second is systematic.}
	\begin{tabular}{@{}
			c
			c
			S[table-format=3.0]@{\,\( \pm \)\,}S[table-format=2.0]
			S[table-format=5.0]@{\,\( \pm \)\,}S[table-format=3.0]
			S[table-format=4.0]@{\,\( \pm \)\,}S[table-format=3.0]
			@{\hspace{1em}(}S[table-format=1.2]@{\,\( \pm \)\,}S[table-format=0.2]@{)$\times 10^5$\hspace{1em}(}
			S[table-format=1.2]@{\,\( \pm \)\,}S[table-format=1.2]@{\,\( \pm \)\,}S[table-format=1.2]@{)$\times 10^8$}}
		\toprule
		& $U_{\textrm{GEM}}$ & \multicolumn{2}{c}{$G_{\textrm{eff}}$} & \multicolumn{2}{c}{$I_{\textrm{ionization}} / e$} & \multicolumn{2}{c}{$\tau$} & \multicolumn{2}{c}{$n_{\textrm{inc}}$} &  \multicolumn{3}{c}{$n_\textrm{tot}$}\\ 
		& (V) & \multicolumn{2}{c}{} & \multicolumn{2}{c}{(\si{\e\per\second})} &  \multicolumn{2}{c}{(s)} & \multicolumn{2}{c}{(\si{\e\per\hole})} & \multicolumn{3}{c}{(\si{\e\per\hole})}\\
		\midrule
		(a) & 400 & 150 & 10 & 84410 & 480 & 6930 & 80  & 5.06 & 0.21 & 2.37 & 0.21 & 0.15  \\[0.5ex]
		(b) & 400 & 150 & 10 & 69330 & 420 & 8200 & 500 & 4.9  & 0.4  & 2.30 & 0.19 & 0.15  \\ \bottomrule
	\end{tabular}
\end{table}

\section{Discussion of the Results}\label{Sec:Discussion}
In the presented measurements, the charging-up effect of a single GEM foil was examined at different X-ray rates and at different GEM voltages.
All measurements show an effective gain which initially increases with time and asymptotically reaches a saturation value.
Each single measurement can be described by an exponential function, see equation~\ref{Eq:Exponential_Gain_Increase}.
This implies that the net accumulation of charges on the polyimide decreases proportionally to the distance from the saturation value, as it is also discussed e.g. in~\cite{Correia_Simulations_2}.
Since the direction of electrons after scattering off gas atoms is almost completely randomized and their mean free path is of the order of a few \si{\micro\meter} \cite{Baille_1981} and hence of similar magnitude as the typical distance to the polyimide, it is to be expected that electrons produced in the vicinity of the insulator inside a GEM hole still may end up at the polyimide also after saturation. This implies that the condition reached after saturation is a dynamical equilibrium rather than a static one, with electrons and ions ending up on the insulator surface at the same rate, instead of being fully repelled by the additional electric field due to the charges on the insulator, as is often conjectured. The result is a constant \emph{net} charge and a constant electric field. 
This is also supported by simulations, as will be reported in a forthcoming paper.

Our measurements show that, for a given GEM voltage, the measured time constants of the charging-up effect are inversely proportional to the rate of incoming electrons.
In order to directly compare measurements performed at different rates,
we switch to the characteristic quantity $n_{\textrm{inc}}$, representing the number of incoming electrons per hole within one time constant $\tau$.
For $U_{\textrm{GEM}}=\SI{400}{\volt}$, four measurements at  different rates of incoming electrons were perfomed.
The measured time constants vary by more than two orders of magnitude, from \SIrange{58}{8200}{\second}. 
When translated to the number of incoming electrons per hole, all measurements agree with each other within uncertainties (see tables~\ref{Tab:Measurement_Method_One_Summary} and  \ref{Tab:Measurement_Method_Two_Summary}. 
From this agreement, one can also conclude that the \emph{charging-down time} (the characteristic time for removal of the charges on the polyimide) is significantly longer than even the longest observed charging-up time. 
A similar conclusion was drawn in~\cite{ALTUNBAS2002177}.

For~$U_{\textrm{GEM}}=\SI{350}{\volt}$, the number of incoming electrons per hole is approximately a factor of ten larger than for $\SI{400}{\volt}$ (see table~\ref{Tab:Measurement_Method_One_Summary}).
This suggests that the total number of electrons in the hole in the characteristic time $\tau$ might be the more appropriate quantity to consider. 
The characteristic total number of electrons in the hole $n_\mathrm{tot}$ can be calculated by multiplying the number of incoming electrons per hole by the effective gain and dividing by the extraction efficiency. 
Using $n_\mathrm{tot}$ as a characteristic quantity reduces the discrepancy between measurements performed at different voltages/gains to a factor of approximately 3.5, but cannot fully reconcile the numbers. According to our measurements, there seems to be a residual dependence of the characteristic charge on the voltage.
Possible reasons for this may be the movement of charges on the insulator surfaces and through the material bulk which may depend on the applied GEM voltage, but also on gas properties like the water content, etc..
Further measurements will be conducted to understand this residual voltage dependence in more detail.

\section{Conclusion and Outlook}
In this work, the charging-up effect of single standard GEM foils was investigated experimentally under well-controlled conditions. In order to study the rate dependence of the effect over a wide range of X-ray interactions in the conversion volume above the GEM foil, we made use of two different measurement methods.
For the first method, a conventional X-ray tube was used which provided a high rate of ionization electrons, allowing for the measurement of currents on the readout anode.
The second method relied on an $^{55}$Fe source that created a comparatively small rate of ionization electrons and on the analysis of $^{55}$Fe spectra taken in short time intervals. 

Our measurements consistently show an initial increase of the effective gain of the GEM foil from the point where the irradiation is started, approaching a saturation value after some characteristic time. Fitting an exponential function to the data, we extracted characteristic time constants and found that they scale inversely proportional to the rate of incoming electrons for a given GEM voltage or gain. For a GEM voltage of~\SI{400}{\volt}, both methods described above could be applied to verify the behavior over a wide range of rates, with characteristic times varying by more than two orders of magnitude. For a lower voltage of~\SI{350}{\volt}, only the X-ray tube could be used as a clean $^{55}$Fe spectrum could not be observed.

In order to compare measurements performed at different X-ray interaction rates and also at different GEM gains, we switch to characteristic quantities describing either the number of incoming electrons per hole or the total number of electrons inside the hole, both within a time span of one time constant.
While both characteristic quantities are found to agree within uncertainties for the two groups of measurements at~\SI{350}{\volt} and~\SI{400}{\volt}, we observe significant differences when comparing measurements at different voltages/gains. 
Assuming that the different gains at different GEM voltages would be taken into account by considering the characteristic total number of electrons inside the GEM hole, we find that there is a residual dependence of this number on the voltage:  $n_\mathrm{tot}=2.4 \times 10^{8}$ for $U_\mathrm{GEM}=\SI{400}{\volt}$, while it is $8.8\times 10^{8}$ for $\SI{350}{\volt}$ (average values).

In the future, we plan to investigate the influence of the GEM voltage on the time constant of the charging-up effect in more detail, also taking into account external parameters like water content. We will also address the question whether the external fields --- drift field and induction field --- also influence the charging-up behaviour.
As previous publications have claimed to observe different signatures for different hole shapes (e.g. single-conical holes), it is interesting to investigate the charging-up effect for these foils as well.
In addition, we are currently performing detailed simulations in order to better understand the microscopic details of the charging-up effect.

\section*{Acknowledgments}
We are indebted to Rui de Oliveira and his team at CERN-EP-DT for providing us with high-quality GEM foils and other detector components. 
We are also grateful to the RD51 collaboration at CERN for interesting and fruitful discussions. 
This work is supported by the German BMBF. 

	\newpage
	\bibliographystyle{unsrt}  


\end{document}